\documentclass[final,5p,times,twocolumn]{elsarticle} 
\usepackage{graphicx}
\usepackage{amsmath}
\usepackage{mathtools}
\usepackage{amssymb}
\usepackage{lineno}
\usepackage[colorinlistoftodos]{todonotes}
\usepackage[colorlinks=true, allcolors=blue]{hyperref}
\journal{Nuclear Instruments and Methods A}

\usepackage{multirow}
\usepackage{multicol}
\usepackage{comment} 

\begin{document}
\begin{frontmatter}

\title{Self-heating Effect in Silicon-Photomultipliers}
\author[A]{E.~Garutti}
\author[A]{R.~Klanner}
\author[B]{E.~Popova} 
\author[A]{S.~Martens}
\author[A]{J.~Schwandt}
\author[A]{C.~Villalba\corref{c}} 

\address[A]{University of Hamburg, Hamburg, Germany}
\address[B]{National Research Nuclear University MEPhI, Moscow, Russia}

\cortext[c]{Corresponding author}
\ead{carmen.villalba@desy.de}

\date{\today}

\graphicspath{ {fig/} }

\begin{abstract}
 
The main effect of radiation damage in a Silicon-Photolumtiplier (SiPM) is a dramatic increase in the dark current. The power dissipated, if not properly cooled, heats the SiPM, whose performance parameters depend on temperature. Heating studies were performed with a KETEK SiPM, glued on an Al$_2$O$_3$ substrate, which is either directly connected to the temperature-controlled chuck of a probe station, or through layers of material with well-known thermal resistance. The SiPM is illuminated by a LED operated in DC-mode. The SiPM current is measured and used to determine the steady-state temperature as a function of power dissipated in the multiplication region of the SiPM and thermal resistance, as well as the time dependencies for heating and cooling. This information can be used to correct the parameters determined for radiation-damaged SiPM for the effects of self-heating. The method can also be employed for packaged SiPMs with unknown thermal contact to a heat sink. The results presented in this paper are preliminary.

\end{abstract}


\end{frontmatter}


\section{Introduction}
The understanding of the response of an SiPM as a function of irradiation fluence is of high relevance for its application in the harsh radiation environment, as at collider experiments.

The aim of this study is to develop a method to determine the SiPM temperature increase induced by the power dissipated in the SiPM amplification layer. This is relevant for the application of SiPMs in high background light, like in LIDAR. Also, in high radiation environments like at the Large Hadron Collider (LHC), where among other effects, there is an increase of several orders of magnitude in the dark current after irradiation \cite{Garutti_2019}. These currents lead to a significant power dissipation, which could produce an increase in the SiPM temperature. 

The performance of the SiPM changes with temperature, $T$. The photo-current at constant bias voltage and constant photon rate decreases with $T$: $I_{photo} \propto  {PDE}(T) \cdot  {G}(T) \cdot {ECF}(T)$ \cite{Klanner_2019}. This is explained by the $T$-dependence of gain ($\mathit{G}$) and photo-detection efficiency ($\mathit{PDE}$) on breakdown voltage $U_{bd}$. An increase in $U_{bd}$ with temperature leads to a decrease in gain and $\mathit{PDE}$:
\begin{align*}
 \mathit{G} & \approx \frac{1}{q_0} \cdot C_{pix} \cdot(U_{bias}-U_{bd}(T)).
\end{align*}

In Ref.~\cite{Lucchini_2020} the authors propose a method to determine the heating of SiPMs for different package configurations and to evaluate the SiPM temperature change 
when operating at high dark count rates. The SiPM current, $I_{SiPM}$, is measured for a constant illumination with light, and the changes in $T_{SiPM}$ are derived from the changes of $I_{SiPM}$, when the bias voltage is switched on, or the thermal conditions changed by switching on a fan. In this paper, we present a similar method to extract temperature variations from current measurements, but $I_{SiPM}$ is measured in a cycle at constant bias voltage and the light intensity is changed. 
A main assumption of the presented analysis is that $I_{SiPM}$ for a constant illumination is only a function of $U_{bias}-U_{bd}(T)$.

The investigated SiPM and the setup are described in Sec.~\ref{sec:setup}. The analysis method and its application to determine the increase in the SiPM temperature is explained in Sec.~\ref{sec:analysis-results}. The main results are given in Sec.~\ref{Sec:results}, followed by the conclusions in Sec.~\ref{sec:Conclusions}.

\section{SiPM investigated and experimental setup} \label{sec:setup}
\begin{figure}[htp]
    \centering
    \includegraphics[width=8cm]{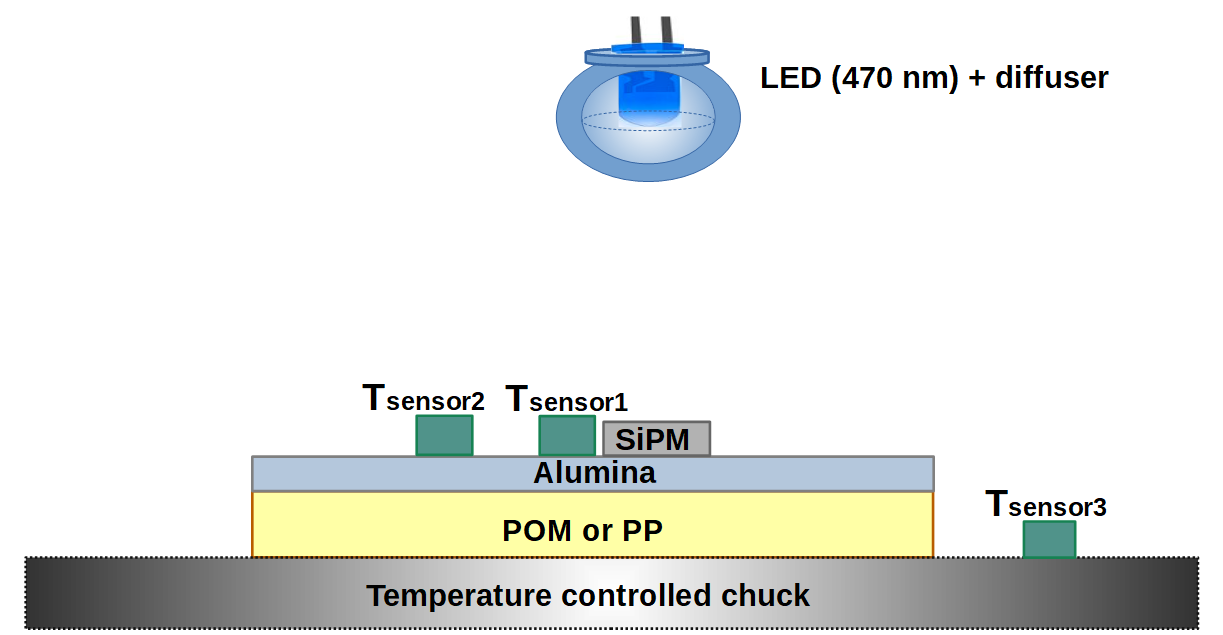}
    \caption{Schematic of the experimental setup. The SiPM is glued on the alumina substrate, with a POM or PP layer to degrade the thermal contact with the chuck. The distances of the $T_{sensors}$ on the alumina from the SiPM center are 3.1 ± 0.1 mm ($T_{sensor1}$) and 7.6 ± 0.1 mm ($T_{sensor2}$).}
    \label{fig:Setup}
\end{figure}

The investigated SiPM is a non-irradiated R\&D SiPM KETEK \cite{ketek} type MP15V09, with $U_{bd}$ = 27.50~±~0.02~V at 25~°C, pixel size of 15×15~$\mu$m$^2$, and number of pixels $N_{pix}$ = 27367. The active region is circular with a radius $r_{SiPM}$ = 1.4~mm. The SiPM is fabricated on a 700~$\mu$m thick silicon wafer. It is glued on an alumina (Al$_2$O$_3$) substrate, with area 20×25~mm$^2$ and thickness 0.6~mm. Two contacts of the SiPM are bonded to the Au traces on the substrate, which are contacted by the probe station needles. The measurements are performed on a temperature-controlled chuck. The alumina substrate is fastened to the chuck using a vacuum pump. For good thermal contact, the substrate is directly placed on the chuck surface. To emulate a degraded thermal contact of the SiPM, two different layers are placed between the alumina substrate and the cold chuck increasing the thermal resistance. One layer 3.1~±~0.1~mm thick of polyoxymethylene (POM) with a thermal conductivity $k$ = 0.3~W/m$\cdot$K, and a polypropylene (PP) thinner layer (1.2~±~0.1~mm) with $k$ = 0.2~W/m$\cdot$K. Three temperature sensors (PT-100) are used, they are glued with a thermal foil, two on the alumina substrate, and one on the chuck; see the sketch of Fig.~\ref{fig:Setup}.
The illumination is provided with light of a wavelength $\lambda$ = 470 nm from a LED operated in DC-mode surrounded by a diffuser. The  setup is located in a light tight and electrically shielded box. For the measurements a Keithley 6517B voltage-source/current-meter 
is used.

Two sets of data are recorded. The calibration data set, with the alumina substrate in direct contact with the chuck, $I_{cal}(U_{cal}, T_{chuck}, I_{LED})$ is taken at several LED currents, $I_{LED}$, with $U_{cal}$ from 0 to 39 V and $T_{chuck}$ from 15 to 35 °C. The second set of data is used to determine the increase in the SiPM temperature from its measured current, in cycles of good and degraded thermal contact. $I_{SiPM}$ is recorded at constant bias voltage, $U_{bias}=38~$V and $T_{chuck} = 25$ °C, for 320~s time intervals with different constant illuminations. This time interval is sufficient for reaching thermal equilibrium. The dark current, $I_{dark}$, is measured when the current of the $I_{LED}$ is zero. The photo-current is defined as $I_{photo}(I_{LED}) = I_{SiPM}(I_{LED}) - I_{SiPM}(I_{dark})$. To dissipate a power around 50~mW in the SiPM the LED current is switched from very low to $I_{LED}$=~0.47~mA. After 320~s the LED is switched off again. $T_{sensor1}$ and $T_{sensor2}$ record the temperature on the surface of the substrate, and $T_{sensor3}$ monitors $T_{chuck}$.

\section{Analysis method} 
\label{sec:analysis-results}

From the calibration data, $I_{cal}(U_{cal}, T_{chuck}, I_{LED})$, the sensitivity, $\mathit{S_{photo}}$ is obtained using Eq. \ref{eq(1)} for different $I_{LED}$ values,
 
\begin{equation} \label{eq(1)}
  \mathit{S_{photo}(U_{bias}, T_{chuck})} = \frac{1}{I_{photo}} \cdot {\frac{\mathrm{d}I_{photo}}{\mathrm{d}U} \cdot {\frac{\mathrm{d}U_{bd}}{\mathrm{d}T}}~ \left[ \frac{1}{\rm{K}} \right]}.
\end{equation}

\begin{figure}[htp]
    \centering
    \includegraphics[width=6cm]{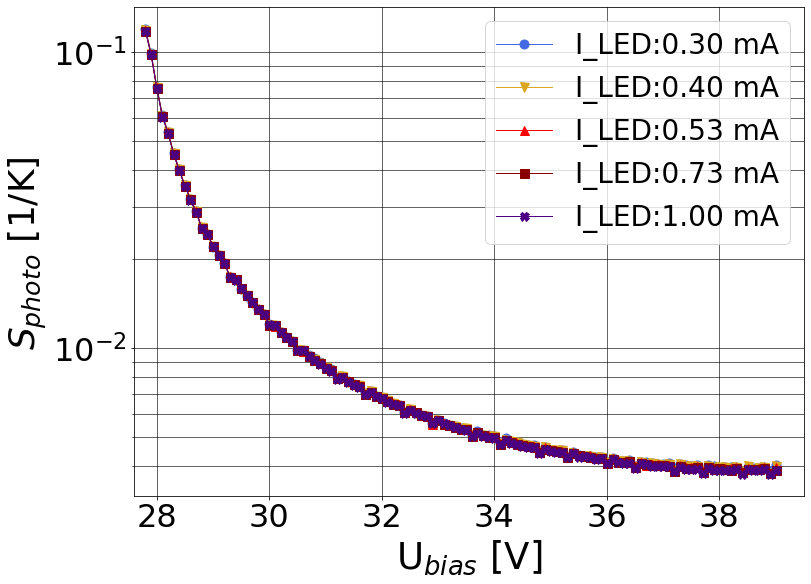}
    \caption{Sesitivity calibration function at several LED currents.}
    \label{fig:Sesitivity}
\end{figure}

The sensitivity gives the relative change of $I_{photo}$ for a temperature change of 1~K. 
Fig. \ref{fig:Sesitivity} shows $\mathit{S_{photo}(U_{cal}, 25~^{\circ} \rm{C})}$ from the calibration data for different LED currents. In the range of interest of the measurements the sensitivity is between 0.004 and 0.01 $ \left[ \frac{1}{\rm{K}} \right]$. It can be seen that $\mathit{S_{photo}}$ does not depend on $I_\mathit{LED}$ showing that heating and occupancy can be ignored for the calibration. 
$I_{photo}$ as a function of the bias voltage increases with $I_{LED}$ for a constant $T_{chuck}$, since the current is proportional to the rate of converted photons producing a Geiger discharge ($R_\gamma$) and the $\mathit{PDE}$,  $~I_{photo} = q_0 \cdot \mathit{G} \cdot \mathit{R}_\gamma \cdot  \mathit{PDE} \cdot  \mathit{ECF}$ \cite{Klanner_2019}. 

\begin{figure}[htp]
    \centering
    \includegraphics[width=7cm]{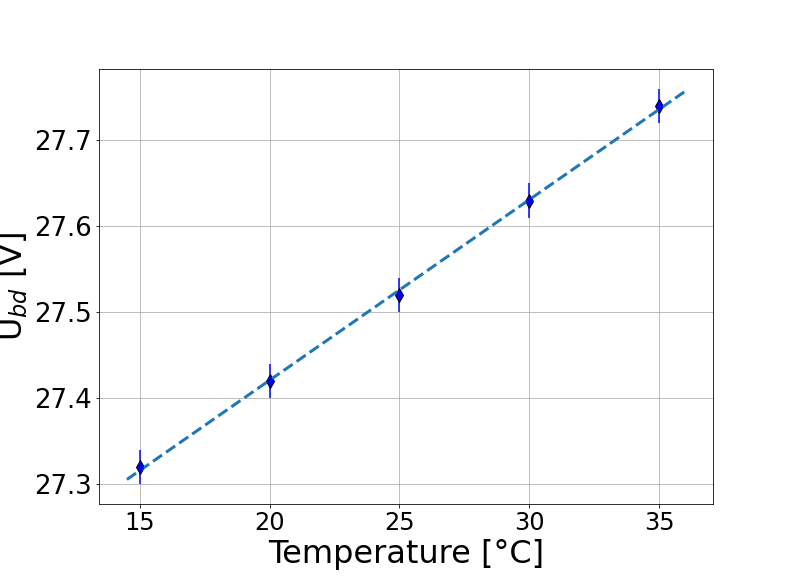}
    \caption{Breakdown voltage as a function of the temperature.}
    \label{fig:VbdvsT}
\end{figure}

The $T$-dependence of the breakdown voltage, shown in Fig.~\ref{fig:VbdvsT}, is derived from the calibration data recorded at different $T_{chuck}$, yielding to $\mathrm{d}U_{bd}/\mathrm{d}T = 21~$mV/K.  

The change in SiPM temperature is calculated, using the sensitivity method, from the ratio of the relative change of $I_{photo}$ divided by the sensitivity, as:

\begin{equation} \label{eq(2)}
 \Delta T_{SiPM} = -\frac{\Delta I _{photo}}{\mathit{S_{photo}} \cdot I_{photo}} .
\end{equation}

\section{Results}
\label{Sec:results}
\subsection{Cycle with good thermal contact}
The first measurement cycle is recorded with the alumina substrate in direct contact with the chuck, $T_{chuck}=$~25 °C. The SiPM is operated at $U_{bias}=38~$V, at t = 640~s the LED is switched to 0.47~mA and $I_{photo}$ rises rapidly as Fig.~4 shows. This LED current results in $I_{photo} \approx 1.2$~mA and a power of 47~mW. 

\begin{figure}[htp]
    \centering
    \includegraphics[width=8cm]{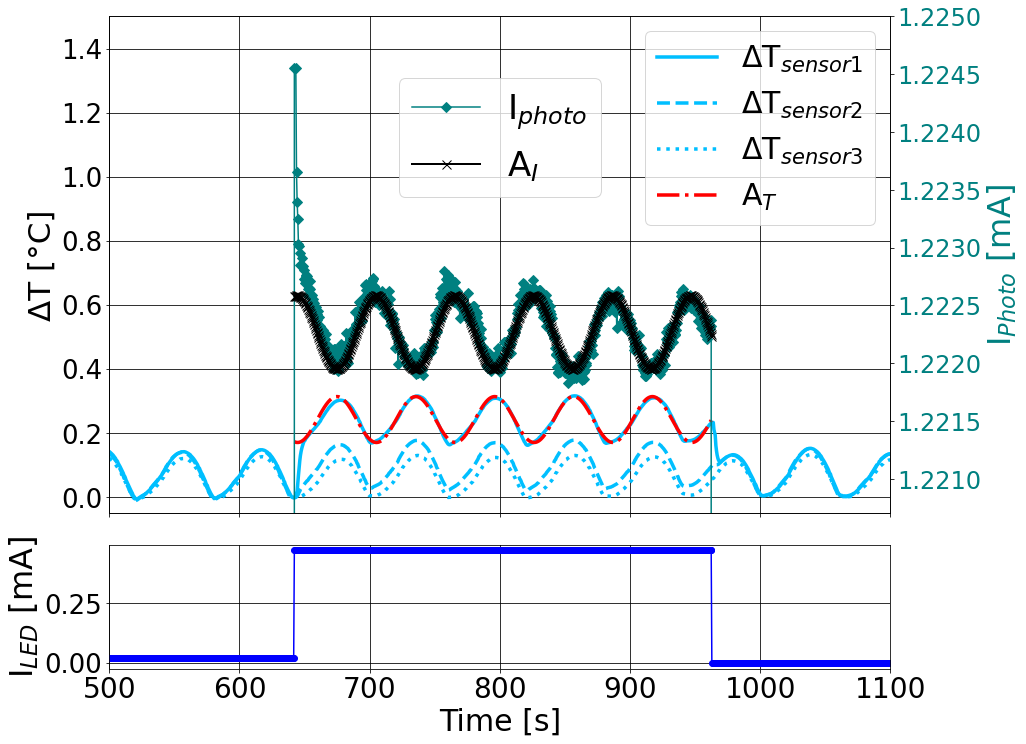}
    \caption{$I_{photo}$ (green), $\Delta T$ of sensors (light blue) and $I_{LED}$ (dark blue) as a function of time for constant $U_{bias}=38~$V and $T_{chuck}=$ 25 °C, for good thermal contact (no layer). It is also included the sinusoidal fit applied to the photo-current (black) and $T_{sensor1}$ (dash-dot red).}
    \label{fig:noPVC_Iph_Tsensors}
\end{figure}

The oscillation of the curves is a consequence of the feedback loop of the temperature-controlled chuck, which stabilizes the temperature to 25.0 ± 0.1~°C. 
All three $T$-sensors have the same amplitude and phase showing that the thermal contact between the chuck and substrate is good.
$I_{photo}$ and $T_{chuck}$ show a 180° phase difference. The reason is that an increase in $T$ increases $U_{bd}$, therefore the gain decreases.

The calculated increase in temperature using Eq.~\ref{eq(2)} is $\Delta T_{SiPM} = 0.56~$K, that corresponds to a breakdown voltage increase $\Delta U_{bd} = 12$~mV. 

\subsection{Cycle with degraded thermal contact}

\begin{figure}[htp]
    \centering
    \includegraphics[width=8cm]{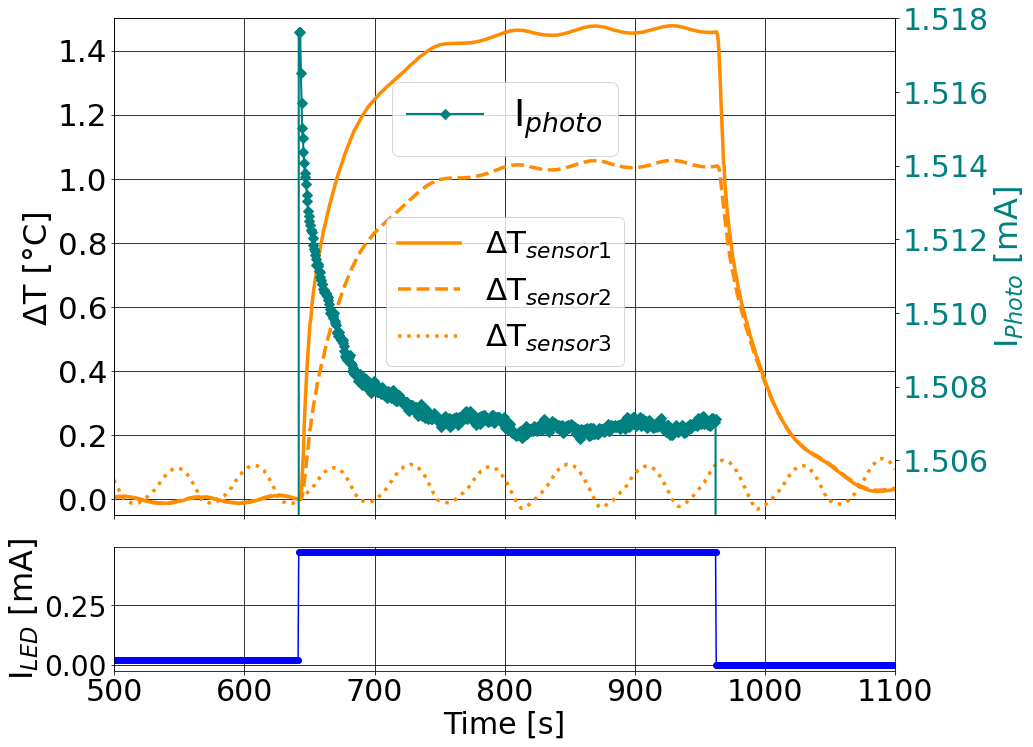}
    \caption{$I_{photo}$ (green), $\Delta T$ of sensors (orange) and $I_{LED}$ (dark blue) as a function of time for constant $U_{bias}=38~$V and $T_{chuck}=$ 25 °C, for degraded thermal contact (3.1 mm POM).}
    \label{fig:PVCthick_Iph_Tsensors}
\end{figure}

Fig.~\ref{fig:PVCthick_Iph_Tsensors} presents the result of the same measurement cycle as discussed in the previous section, however in this case with the POM layer between the chuck and the alumina substrate. It can be observed that it takes about 60 s until the steady state is reached. From the change of $I_{photo}$, $\Delta T_{SiPM} = 1.87~$K is obtained using Eq.~\ref{eq(2)}, which corresponds to $\Delta U_{bd} =39~$mV. Because of the 3.1 mm POM layer the distance between the LED, that is fixed at the same position, and the SiPM is reduced, therefore the photon flux is increased, $I_{photo}$~$\approx$1.5~mA, and the dissipated power in the SiPM is 58~mW.

It can also be seen that there are phase differences between the $T$-oscillations of $T_{sensor1}$ and $T_{sensor2}$ relative to $T_{chuck}$, and that $\Delta T$ decreases with the distance between sensor and SiPM.

\subsection{Cross-check}

In order to cross-check the values of $\Delta T_{SiPM}$ obtained from the sensitivity method, the temperature-controlled chuck oscillations were exploited. 
Based on the discussion of the previous section, it can be assumed that the amplitudes of $T_{SiPM}$ and $T_{sensor1}$ oscillations are approximately the same. A sinusoidal fit was applied to determine the amplitude of the oscillation for $T_{sensor1}(t)$ and $I_{photo}(t)$ for the data with good thermal contact, and from the ratio is obtained the calibration constant:
\begin{equation} \label{eq(3)}
 \alpha = \frac{A_T}{A_I}  = 0.23~ \mathrm {\left[ \frac{K}{\mu A} \right]},
\end{equation} this allows to estimate the change in SiPM temperature using: 
\begin{equation} \label{eq(4)}
    \Delta T_{SiPM} = - \alpha \Delta I_{photo}.
\end{equation} For the case of good thermal contact and a power of 47 mW, which is the same data used for the sensitivity method, a $\Delta T_{SiPM} =$ 0.62~K is obtained, which is 10~\% larger than the value from Eq.~\ref{eq(2)}. 

For the data with the POM and the PP layer between the SiPM mounted on the alumina substrate and the cold chuck, the current is normalized to the maximum value of the data with good thermal contact, and using alpha from Eq.~\ref{eq(3)} the increase in SiPM temperature for degraded thermal contact is estimated with Eq.~\ref{eq(4)}. All the results are summarised in Tab.~\ref{table:1}. 

\subsection{Summary of results}
\begin{figure}[!h]
    \centering
    \includegraphics[width=7cm]{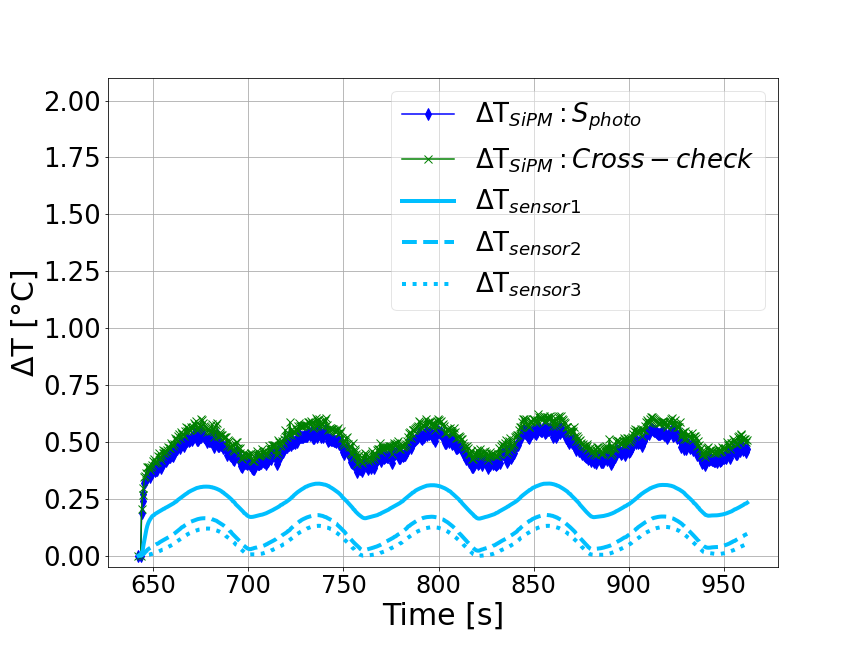}
    
    (a)
    \vspace{1.10mm}
    
    \includegraphics[width=7cm]{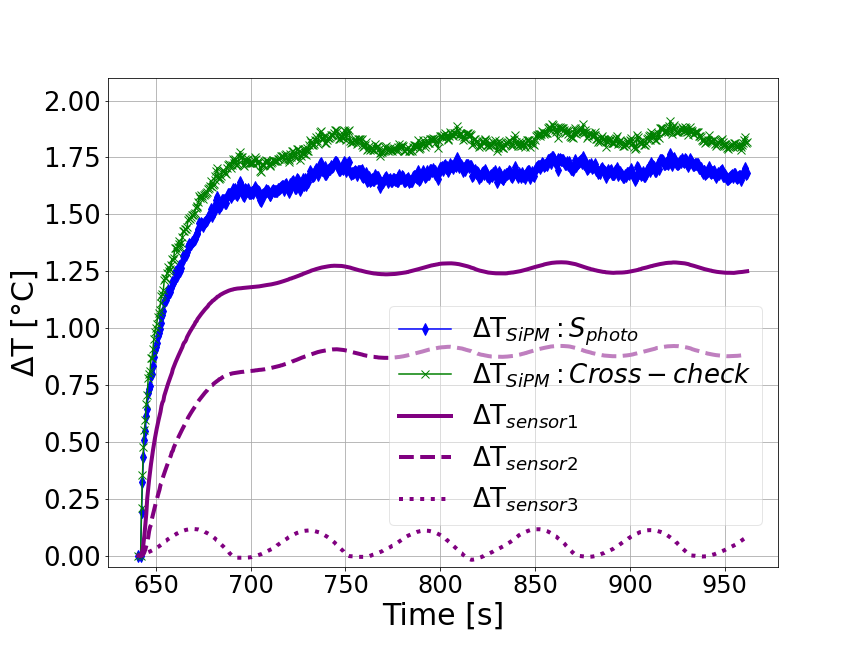}
     
    (b)
    \vspace{1.10mm}
    
    \includegraphics[width=7cm]{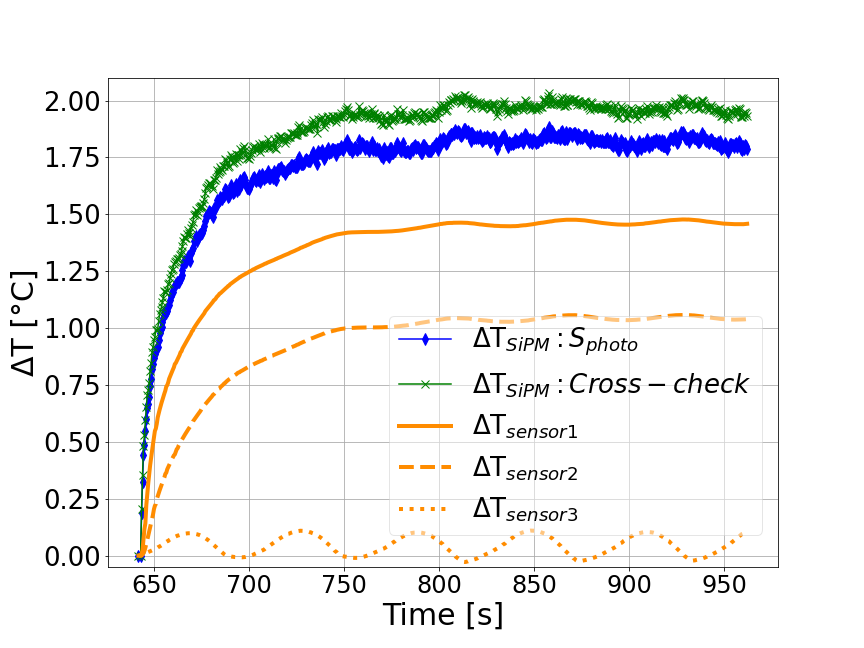}
    
    (c)

\caption{Determined $\Delta T_{SiPM}$ using the sensitivity method (blue) and the cross-check (green) as a function of time for the interval of maximum power dissipation. Results for no layer (a), a 1.2 mm thick PP layer (b), and a 3.1 mm thick POM layer (c). }
\label{fig:deltaT_comparison}
\end{figure}

Fig. \ref{fig:deltaT_comparison} presents the summary of the  results obtained; Fig.~6~(a) for good thermal contact, (b) for a 1.2 mm PP layer, and (c) for a 3.1 mm POM layer. The $\Delta T_{SiPM}$ using the sensitivity method, Eq.~\ref{eq(2)}, as a function of time is plotted in blue and compared to the values obtained with the method of Eq.~\ref{eq(4)}, in green. The cross-check values are $8$ to $10\,\%$ larger. The temperature increase with a POM or PP layer is larger than without any layer, because the heat produced in the SiPM is not efficiently transported through the insulating material.
\begin{table}[htp]
\footnotesize
\centering
\begin{tabular}{|p{0.9cm}|p{0.8cm}|p{1.2cm}|p{1.4cm}|p{1.cm}|p{0.9cm}|} 
    
    \hline & $\mathbf{P}[\mathrm{mW}]$ & $\Delta \mathrm{T}_{\text {SiPM }}[\mathrm{K}]$ & Cross-check $\Delta \mathrm{T}_{\text {SiPM }}[\mathrm{K}]$ & Rel. difference\\
    \hline \hfil no~layer & \hfil $46.55$ & \hfil $0.56$ & \hfil $0.62$ & \hfil $10.3 \%$ \\
    \hline \hfil  PP & \hfil $50.81$ & \hfil $1.74$ & \hfil $1.91$ & \hfil $9.5 \%$\\ 
    \hline \hfil POM & \hfil $57.68$ & \hfil $1.87$ & \hfil $2.03$ & \hfil$8.6 \%$\\ 
    \hline

\end{tabular}
\caption{Calculated $\Delta T_{SiPM}$ for each thermal resistivity using the sensitivity method and the cross-check $\alpha$.}
\label{table:1}
\end{table}

\begin{table}[htp]
\footnotesize
\centering
\begin{tabular}{|c|c|c|}
\hline & $\mathit{d}~[\mathrm{mm}]$ & $\mathit{k}~[\mathrm{W/m\cdot~K}]$ \\
    \hline \hfil PP & \hfil  $1.2$  & \hfil $0.2$ \\
    \hline \hfil  POM & \hfil $3.1$  & \hfil $0.3$ \\
    \hline

\end{tabular}
\caption{ Thickness ($d$) and thermal conductivity ($k$) for the polypropylene (PP) and the polyoxymethylene (POM) layer.}
\label{table:2}
\end{table}

\section{Conclusions} \label{sec:Conclusions}

A method is developed to determine the self-heating of a SiPM from the measurement of its photo-current. Self-heating causes an increase in breakdown voltage and a decrease in SiPM current. 
In our measurements, the SiPM is illuminated by a blue LED and the SiPM current is measured at constant bias voltage when changing the LED current.

The developed method is used to determine the temperature increase for a SiPM with dissipated power around 50 mW. 
This is the measured power of a Hamamatsu SiPM (MPPC S14160-976X) irradiated to a fluence $\Phi_{eq}= 10^{13}~\rm{cm}^{-2}$ and operated at 2~V above breakdown voltage at $- 30\,^\circ$C.
For different thermal resistances between the SiPM and the chuck, the following results are obtained: for good thermal contact an increase of about 0.5 K; for degraded thermal contact, similar to that of a SiPM mounted on a PCB, the increase in temperature is about 2 K, corresponding to an increase in $U_{bd}$ of 40 mV. 
A simple cross-check confirmed within a 10~\% the temperature increase obtained using this calibration method. 

Next, the method will be extended by determining $T_{SiPM}$ using $I_{dark}$ during the cool down phase after switching off the illumination, and using measurement cycles changing the dissipated power. Finally the heating of irradiated SiPMs will be investigated. Last, but not least it is planned to compare the measured time dependence $T_{SiPM}(t)$ to thermal simulations.

\section*{Acknowledgement}
  
We acknowledge the support from BMBF via the High-D consortium.
This work is supported by the Deutsche Forschungsgemeinschaft (DFG, German Research Foundation) under Germany's Excellence Strategy, EXC 2121, Quantum Universe (390833306).

\bibliographystyle{elsarticle-num}
\bibliography{article}

\end{document}